# Neutrino Spectra and Uncertainties for MINOS

Sacha Kopp

*Department of Physics, University of Texas, 1 University Station C1600*
*Austin, Texas, 78712 U.S.A.*

**Abstract.** The MINOS experiment [1] at Fermilab has released an updated result on muon disappearance [2]. The experiment utilizes the intense source of muon neutrinos provided by the NuMI beam line. This note summarizes the systematic uncertainties in the experiment's knowledge of the flux and energy spectrum of the neutrinos from NuMI.

## INTRODUCTION

MINOS [1] studies the disappearance of muon neutrinos across 734 km, comparing their energy spectrum at two detectors sited at either end of this flight path. The first, "Near Detector" (ND) is sited at Fermilab, and the "Far Detector" (FD) is located in the Soudan Mine. Updated results on $\nu_\mu$ disappearance and the inferred neutrino oscillation parameters were presented by Žarko Pavlović at this conference [2].

The MINOS experiment relies upon very precise knowledge of the expected energy spectrum of $\nu_\mu$ arriving at the FD in the absence of oscillations. In part, small systematic uncertainties are achieved by utilizing a $\nu_\mu$ beam derived from the NuMI beam line, whose spectrum can be controlled. In part, such low systematic uncertainties are achieved by having two detectors, the first of which (the ND) measures directly the spectrum in the absence of oscillations. Because MINOS is a two-detector experiment, the uncertainties for the oscillation measurement are best characterized by the uncertainty in the ratio of FD to ND fluxes, "F/N". Figure 1 compares the $\nu_\mu$ spectrum observed in the ND to our beam MC.

The experiment relies upon two analysis strategies to reduce systematic uncertainties from the beam spectrum. The first is a set of beam data used to measure or constrain uncertainties due to particle focusing in the beam line. The second is a flexible beam design which allows us to manipulate the spectrum in known fashion so as to infer directly the production of particles off the NuMI target, rather than relying solely upon external measurements. Each of these analyses are discussed in turn.

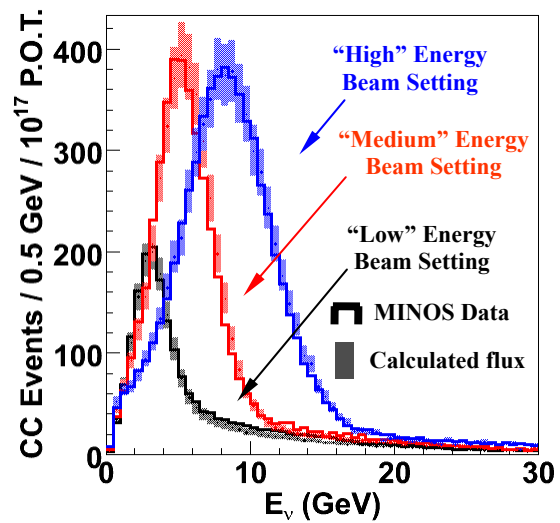

**FIGURE 1.** Energy spectra of charged-current (CC) $\nu_\mu$ interactions in the MINOS Near Detector under three neutrino beam configurations. The MINOS data is compared to the NuMI flux calculation [1], which relies on the FLUKA2005 [3] cascade Monte Carlo to predict the flux of hadrons off the target.

## FOCUSING UNCERTAINTIES

The Main Injector 120 GeV proton beam is impinged upon the NuMI graphite target for production of $\pi$ and $K$ secondaries. These secondaries are sign- and momentum-selected by a pair of focusing horns [4]. To correctly predict the neutrino spectrum it is important to understand precisely particle tracking through the horns. We rely on a series of dedicated beam studies to test this tracking.

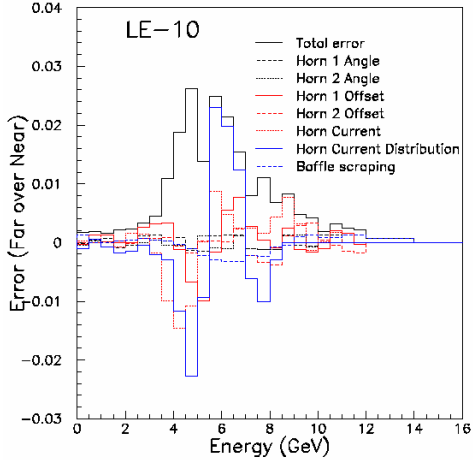

**FIGURE 2.** The relative uncertainty in the ratio of FD and ND neutrino energy spectra due to beam focusing effects. Contributions to the overall error are described in the text.

Alignment of the target and horns with respect to the beam axis was accomplished with a set of dedicated runs in which the proton beam was scanned transversely across these components [5]. Remaining uncertainties in their locations is ±(0.5-1.0)mm.

The magnetic field in the horns was measured prior to their installation in the target station. The remaining uncertainties in the field arises from the ~1 kA uncertainty in the current flowing in the horns, and the uncertainty in the distribution of the current through the horns' conductor, whose 3 mm thickness is comparable to the ~7 mm skin depth expected at the frequency of the current pulse.

The largest focusing uncertainties in the ratio of spectra in the FD and ND are summarized in Figure 2. In the focusing peak, these uncertainties are negligible, and at the edge of the peak they are 2-3%.

## TARGET YIELD UNCERTAINTIES

In the limit of perfect horn focusing, the exact production spectra of secondary mesons off the NuMI target does not contribute to the uncertainty in the ratio of FD and ND fluxes. However, because of the finite acceptance of the beam line, imperfections in the focusing, and the lack of focusing for very forward particles off the target which pass through the gaps in the horns, some knowledge of the production spectra vs $(x_F, p_T)$ is necessary. We have utilized the Fluka2005 cascade Monte Carlo [3] to estimate these target yields.

As may be seen in Figure 1, the Fluka yields, together with our particle tracking Monte Carlo, describes the MINOS ND reasonably well. Several configurations of the NuMI target and horns were operated, and data in the ND accumulated. Discrepancies between the data and MC are within the 30-40% span in predictions [3,6-8].

To reduce the uncertainty in prediction of the FD spectrum (see Figure 3), we constrain the poorly-known yield of secondaries off the NuMI target using the ND data. The NuMI target can be located at several positions upstream of the horns, thereby altering the $(x_F, p_T)$ of hadrons off the target which are focused by the horns [9] and altering the expected neutrino spectra in the MINOS detectors. By acquiring data in the ND under several such configurations (see Figure 4), it is possible to, in part, deconvolve discrepancies which arise from neutrino cross section errors, from those arising from focusing errors, to those arising from incorrect knowledge of particle yields off the target at a given $(x_F, p_T)$.

We developed a simple parametric function [10] to describe the yield of target secondaries as a function of $(x_F, p_T)$ which is based upon the function of Ref. [8]. The parameters of this function were varied to give the best fit to the ND data, and the resulting distributions are shown in Figure 4. The target yields obtained from this tuning procedure then are used to calculate a better prediction for the F/N ratio, as shown in Figure 3. The remaining uncertainty in the parameters of this function result in a (0.5-5.0)% uncertainty in the F/N ratio, as shown in Figure 3. This uncertainty is approximately half the uncertainty prior to the tuning.

The MINOS detectors permit separation of $\nu_\mu$ and $\bar{\nu}_\mu$, the latter of which arise from $\pi^-$ and $K^-$ off the target, while the $\nu_\mu$ arise from $\pi^+/K^+$. By fitting the $\nu_\mu$ and $\bar{\nu}_\mu$ spectra in the ND, we estimate the ratio of yields $\pi^+/\pi^-$ off the target (see Figure 5) which is in good agreement with measurements from NA49 [11].

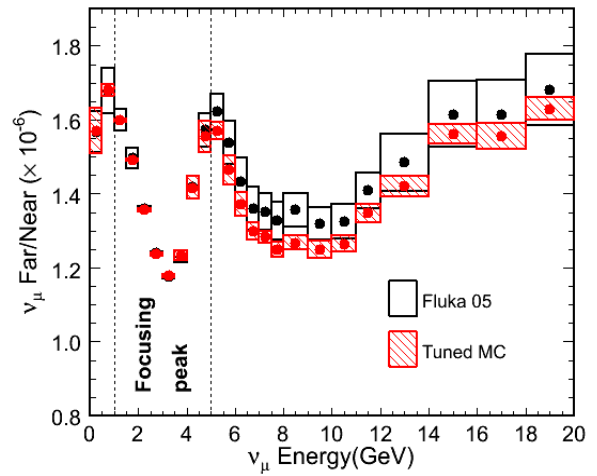

**FIGURE 4.** The ratio of the expected FD to ND neutrino flux, in the absence of oscillations, as a function of energy. Shown is the Monte Carlo prediction using Fluka [3] (open boxes) and the target yields tuned to agree with the ND data (shaded boxes). Box sizes indicate uncertainties.

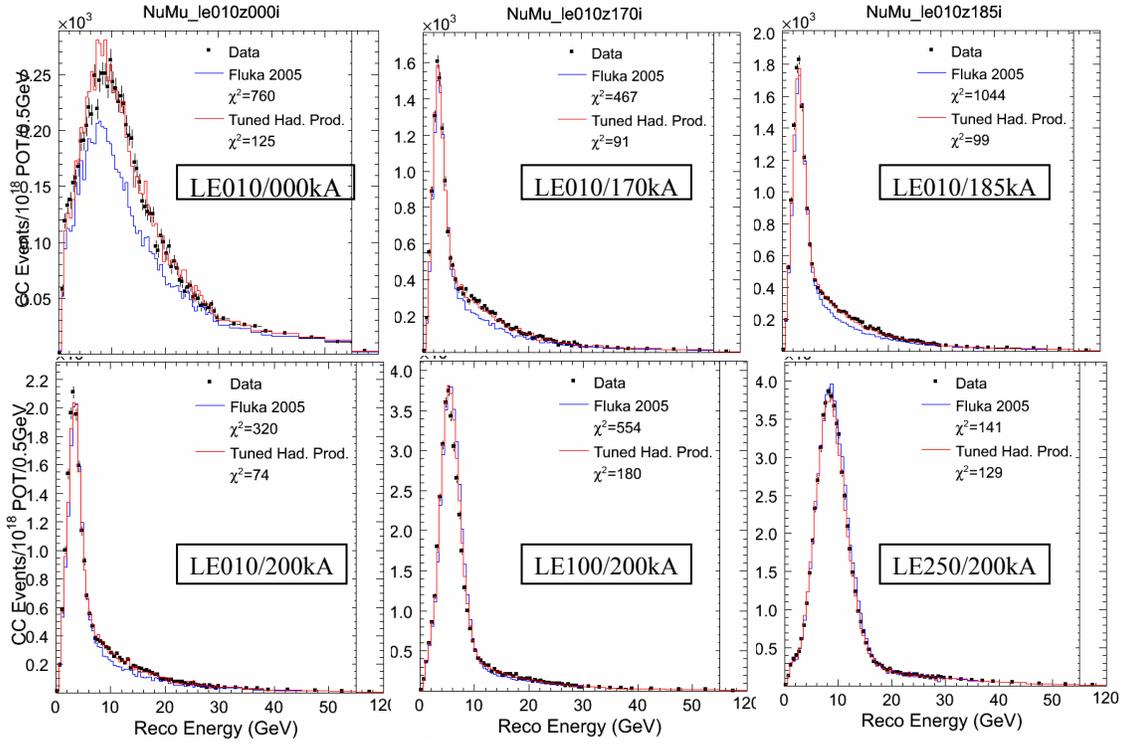

**FIGURE 3.** Energy spectra of charged-current $\nu_\mu$ interactions in the MINOS ND under six different neutrino beam configurations [9] of horn current and target position. Adjustment of the $(x_F, p_T)$ of secondaries off the target brings the calculated spectra into agreement with the MINOS data. The MINOS data are the points, the blue curves are the flux predictions using the FLUKA [3], and the red curves are the result of a fit to the secondaries' yields to the MINOS data.

The improved constraints on the yields of secondary mesons off the target is also of benefit for the $\nu_\mu \to \nu_e$ search at MINOS. The flux of $\nu_e$ which is intrinsic in the beam, arising predominantly from muon decay (where the muons are produced in pion decay) and $K_{e3}$ decays, comprise a background for this search. The uncertainties in pion and kaon yields off the target thus contribute a background uncertainty for the $\nu_\mu \to \nu_e$ measurement. The constraints on these secondary yields provided by the $\nu_\mu$, however, limit the uncertainty on the instrinsic background to ~6%, as is discussed by Mayly Sanchez at this workshop [12].

## REFERENCES

1. D.G. Michael *et al*, Phys. Rev. Lett. **97**:191801, 2006
2. Z. Pavlović, presented at this workshop
3. A. Ferrari, P. Sala, A. Fasso, J. Ranft, CERN-2005-010.
4. A. Abramov *et al*, Nucl. Instr. Meth. **A485**, 209 (2002).
5. R. Zwaska *et al*, Nucl. Instr. Meth. **A568**:548, 2006
6. R. Brun *et al*., GEANT, CERN Program Library Long Writeup W5013 (1994).
7. N. Mokhov *et al*, FERMILAB-CONF-03-053, 2003.
8. M. Bonesini *et al*, Eur. Phys. J. **C20**, 13, 2001.
9. M. Kostin *et al*, FERMILAB-TM-2353-AD, 2001.
10. D.G. Michael *et al*, FERMILAB-PUB-07-577-E (submitted to Phys. Rev. **D**).
11. C. Alt *et al*, Eur.Phys. J.**C49**, 897, 2007
12. Mayly Sanchez, presented at this workshop.

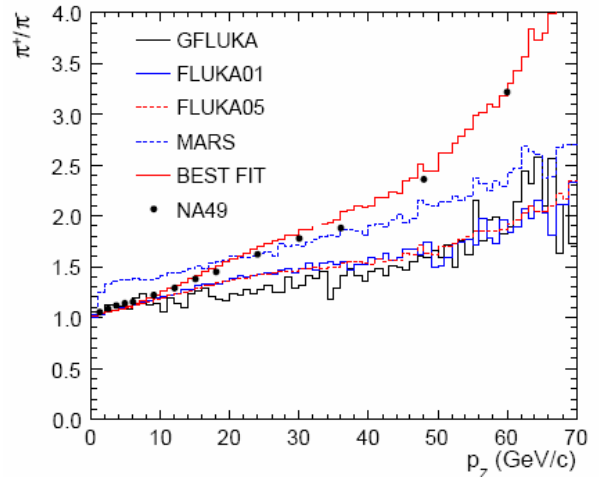

**FIGURE 5.** The ratio positive to negative pions off the target as predicted by several Monte Carlo codes and by our fit to the MINOS ND data. The MINOS fit is compared to measurements from NA49 [11].